\input phyzzx
\sequentialequations
\overfullrule=0pt
\tolerance=5000
\nopubblock
\twelvepoint

\line{\hfill }
\line{\hfill IASSNS 95/110}
\line{\hfill cond-mat/9512156}
\line{\hfill December 1995}
\line{\hfill DRAFT}

\REF\paperone{C. Nayak, F. Wilczek {\it 
Spin-Singlet Ordering Suggested by Repulsive
Interactions}, IASSNS 95/75, PUPT 1570, cond-mat/9510132.}

\REF\papertwo{C. Nayak, F. Wilczek {\it 
$Z_2$ Symmetry Breaking for the Mott Insulator
Phenomenon}, IASSNS 95/111, in preparation.}
 
\REF\polyA{A. Heeger, S. Kivelson, J. R. Schrieffer, and W.-P. Su 
{\it Rev. Mod. Phys}. {\bf 60}, 781 (1988); R. Jackiw, C. Rebbi
{\it Phys. Rev}. {\bf D13}, 3398 (1976).}

\REF\slacbag{W. Bardeen, M. Chanowitz, S. Drell, M. Weinstein, and 
T.-M. Yan {\it Phys. Rev}.  {\bf D11}, 1094 (1975).}

\REF\helium{T. L. Ho, J. Fulco, J. R. Schrieffer, and F. Wilczek
{\it Phys. Rev. Lett}. {\bf 52}, 1524 (1984).}

\REF\volovik{G. Volovik {\it Is There Analogy Between Quantized Vortex and 
Black Hole?} , Helsinki preprint LTL-95006 (1995).}

\titlepage
\title{Aspects of d-Density Order\foot{Invited Talk at the
Pacific Conference on Condensed Matter Theory:  Complex Materials and
Strongly Correlated Systems, Seoul Korea December 2-5, 1995.}}

\author{Frank Wilczek\foot{Research supported in part by DOE grant
DE-FG02-90ER40542.~~~wilczek@sns.ias.edu}}
\vskip.2cm
\centerline{{\it School of Natural Sciences}}

\centerline{{\it Institute for Advanced Study}}
\centerline{{\it Olden Lane}}
\centerline{{\it Princeton, N.J. 08540}}
 
\endpage
 
\abstract{I briefly review the concept of d-density ordering,
extend it to arbitrary dimensions, and speculate that it might
describe Mott insulators.  
This ordering supports zero modes on domain
walls, and quite plausibly dopants occupy such states.  
This phenomenon could induce quasi-one
dimensional behavior in a two-dimensional electron system.  }
 
\endpage

In this talk I will briefly describe some recent and ongoing work done
in collaboration with Chetan Nayak [\paperone , \papertwo].  

\chapter{Introduction}

One has become accustomed to the existence of subtle
ordering in particle-particle channels, especially 
in connection with the superfluidity of $^3$He, and also in the
apparent
d-wave superconductivity of cuprates and the less certainly determined,
but apparently non-s wave, pairings for heavy fermion systems.  On the
other hand several familiar types of ordering, including spin-density
wave (and thus ferromagnetic and antiferromagnetic, among others) and
charge-density wave, can be described as pairing in particle-hole
channels.
It is therefore interesting to inquire whether orderings involving
more subtle
particle-hole pairings occur naturally -- in models, and in reality --
and if so, whether they might have important physical consequences.  

Here I will discuss a particular example of such ordering, which for
reasons that will become apparent we call d-density ordering.  I will
show that it appears very naturally in a simple model of considerable
physical interest, and that both the ground state and the
quasiparticles have quite interesting properties, suggesting that this
type of ordering may have something to say about the Mott insulator
phenomenon.  In this connection it is particularly
interesting that the quasiparticles plausibly are localized on
domain walls.  That association 
provides a natural mechanism whereby quasi-one
dimensional dynamics could govern interactions 
among the quasiparticles in a two dimensional system.

\chapter{Ground State; Symmetry Properties}

In [\paperone ] we arrived at the possibility of a new kind of
ordering
by studying a particular model Hamiltonian; here  I
shall instead first define the ordering in a general way, and only then
discuss its dynamical origin. 
Consider a two-dimensional square lattice with spacing $a$, aligned
along the coordinate axes.    
It is
natural, for several reasons, to consider symmetry breaking at
momentum
$G/2 \equiv ({\pi\over a}, {\pi\over a})$.  Indeed, ordering at such a
momentum -- half a reciprocal lattice vector -- can be induced by 
quartic interactions at the mean-field level, as we shall discuss
concretely in a moment.  Well-established possibilities, including
antiferromagnetic and commensurate
charge-density order are described in this way.  Also in two
dimensions this momentum vector has a special significance, since the
tight-binding Fermi surface nests here at half filling.  The d-density
order is characterized abstractly by the expectation value
$$
\langle c^{\dagger\alpha}_{k+G/2} c_{\beta{k}} \rangle ~=~ 
i \delta^\alpha_\beta f(k)~. 
\eqn\order
$$
Here $\alpha , \beta$ are spin indices, and $f$ is a real function
that changes sign upon rotation by $\pi/2$.  The ordering defined by
\order\ breaks three important symmetries:
\bigskip
$\bullet$ Translation by one lattice spacing (because of the momentum transfer 
$G/2$).
\bigskip
$\bullet$ Time reversal symmetry, as suggested by the appearance of i.
\bigskip
$\bullet$ The discrete rotation through $\pi/2$. 
\bigskip
However the square of these operations, or any  product of any two
of them, {\it is\/} a valid symmetry.  There is no macroscopic T
violation,
therefore.  The broken symmetry is
parametrized by the simplest of all groups, $Z_2$.  

Because one obtains the true density by integrating \order\ over $k$
there
is a cancelation, and the charge distribution is homogeneous.
However the ordering in its mathematical form does resemble a d-wave
analogue of a charge density wave, which explains the name.
There is also the possibility of d-spin density wave ordering, but I
will not discuss that further here.  Our d-density order parameter, of course,
is
a spin singlet.

One can also discuss the ordering in real-space terms.  The relevant
vacuum expectation value is of the form
$$
{\rm Im }~ \langle c^{\dagger\alpha} (x+a\hat n) c_\beta (x) \rangle
~=~ 
\delta^\alpha_\beta g(x)
\eqn\spaceorder
$$
where $g$ is a real function that changes sign upon translation by a 
minimal lattice vector, and $\hat n $ is a unit vector along a
coordinate axis.   In this form we see some similarity between
d-density
ordering and staggered flux phases.  In fact microscopic currents do flow in
the d-density ground state; however  their value does not correspond
to any simple value of the flux ({\it e.g}. 0, $\pm \pi/2$,  $\pi$).

Upon d-density ordering the effective size of the unit cell is
doubled, and thus a gap opens at half filling.  The Mott insulator
phenomenon, in its most basic form, 
is therefore an automatic consequence.

For definiteness I have in the preceding supposed 2-dimensional
kinematics, however let me emphasize that the concepts extend in a
most natural way to other dimensions.  Above, I have taken 
care to
formulate them in such a way that the extension is obvious (this may
not have been transparent in [\paperone ]).     

To judge the physical reasonableness of d-density order we must
consider whether it arises in the ground state of some simple,
significant Hamiltonian.  

Though in some sense it is a case of 
``old wine in new
bottles'', we think it is quite noteworthy and remarkable that the
1-dimensional version arise for the very simplest Hubbard model at
half filling.  This is most easily seen upon bosonizing the theory.
The analysis is detailed in [\papertwo ]; here I shall be very
telegraphic.
The Fermi surface splits into two points, describing left- and
right-movers.
Spin and charge excitations travel at different velocities, and are
conveniently represented by different fields.  Let the charge field
for left and right movers be represented by the appropriate
components $\chi_{L, R}$ of a scalar
field $\chi$ compactified on a circle of radius $1/\sqrt 2$.  At generic
values
of the filling, these are free fields.  Precisely
at half filling there is an additional relevant interaction, the 
Umklapp process, that in
terms of the electron fields is
$$
{\cal L}_{\rm umklapp} ~=~ 
u(\epsilon_{\alpha\beta} \psi^{\dagger\alpha}_L \psi^{\dagger\beta}_L )
 (\epsilon^{\gamma\delta} \psi_{R\gamma} \psi_{R\delta} )
\eqn\electumk
$$
and for the holon fields induces the simple form
$$
{\cal L} ~=~ {1\over 2} (\partial\chi)^2 ~+~ 2u \cos \sqrt 2 \chi~.
\eqn\holonumk
$$ 
Here $\chi \equiv \chi_L + \chi_R$ and $u = {U\over 16t}$, where $t$ is
the
hopping parameter.  This holon Lagrangian is invariant under the $Z_2$
symmetry $\chi \rightarrow -\chi$ which interchanges the two Fermi
points.  For an attractive interaction $u < 0$ the energy is minimized
at
$\langle \chi \rangle = 0$, and for a repulsive interaction $u > 0$ the
energy is minimized at $\langle \chi \rangle = \pi/\sqrt 2 $.  In the
latter
case the $Z_2$ symmetry is broken.

In terms of the holon variables
$\psi_L \equiv e^{-i\chi/\sqrt 2}$, 
$\psi_R \equiv e^{i\chi_R/\sqrt 2}$, $\langle \chi \rangle = 0 $
corresponds
to 
$$
\langle \psi_{R} \psi^{\dagger}_L \rangle ~=~
\langle \psi_{L} \psi^{\dagger}_R \rangle ~=~ c~,
$$
a real number.  This is a conventional, commensurate charge density
wave.  On the other hand $\langle \chi \rangle = \pi/\sqrt 2$
corresponds to
$$
\langle \psi_{R} \psi^{\dagger}_L \rangle ~=~
-\langle \psi_{L} \psi^{\dagger}_R \rangle ~=~ if ~,
\eqn\onedeorder
$$
a pure imaginary number.  As a result of the minus sign this state
does not have charge-density order, contributions from L-R and 
R-L terms canceling.  Instead, \onedeorder\ is precisely of the
d-density form.


In [\paperone ] we analyzed a two-dimensional generalized 
Hubbard model with interaction
Hamiltonian
$$
{\cal H}_{\rm int.} ~=~ 
U\sum_x c^\dagger_\uparrow (x) c_\uparrow (x)  
   c^\dagger_\downarrow (x)c_\downarrow (x)
+ {1\over 2} V\sum_{x, x^\prime} c^{\dagger\alpha} (x)c_\alpha (x)
  c^{\dagger\beta} (x^\prime )c_\beta (x^\prime)
\eqn\modelHam
$$
containing both on-site and nearest neighbor repulsion.  We found that
d-density order is plausibly the ground state for a range of $U$ and
$V$ around $U \approx 4V$ near half filling.  This was done by
considering
BCS-like trial wave functions of the form 
$$
\Psi ~=~ \prod_k \, (u_{k\uparrow} c^\dagger_{k\uparrow}
+ v_{k\uparrow} c^\dagger_{k+G/2 \uparrow} )\,
(u_{k\downarrow} c^\dagger_{k\downarrow}
+ v_{k\downarrow} c^\dagger_{k+G/2 \downarrow} )\, |0 > ~,
\eqn\trialfunction
$$
and minimizing the energy as a function of the $u_k, v_k$ (subject of
course
to
$|u_k|^2 + |v_k|^2 = 1$).   The interaction energy that drives one
toward  d-density order arises entirely from the $V$ term.  Note that
away from perfect nesting the transition occurs at a finite coupling.

\chapter{Quasiparticles on Domain Walls?}

In the previous section, I have described a form of $Z_2$ symmetry
breaking that opens up a gap at half filling.   With a $Z_2$ symmetry
breaking, there is the possibility of domain walls marking the
boundary between the different phases.  In different 
languages and in various contexts (see for example 
[\polyA , \slacbag , \helium , \volovik ]), 
it is known that in this kind of situation
there  is a very interesting dynamical possibility associated with the
domain walls.  The particles which acquire mass from the transition
will have a positive mass on one side and a negative mass on the
other.
Normally of course in a homogeneous phase we can define the negative
mass away, but here the relative sign is important.  It leads to the
formation of zero-modes.  In the present context, this means that
there are mid-gap excitations localized to such walls.  

It is quite conceivable that in the presence of d-density order
the most favorable way to accommodate
electrons doped away from half filling is to associate them with
pre-existing domain walls, or even that such particles spontaneously
nucleate wall-bits (very much in the spirit of the ``SLAC bag''
[\slacbag ]).   In one dimension this is the case: charged excitations
are
associated with solitons of the sine-Gordon model \holonumk . 

Whenever it occurs, the dynamical consequences of localizing 
dopants to domain walls are sure to
be profound.  For example, in two dimensions the dopants would find
themselves confined to one-dimensional lines.  In this situation,
their interactions generically induce non-Fermi liquid behavior,
including
spin-charge separation.  Also, the walls  must be treated as 
dynamical objects -- for example, even if a long wall
is pinned by impurities at
several locations it will be able to wiggle in between, and there will
be an effective ``electron-phonon'' interaction of quite unusual type
between its wiggles and the dopants.   It may be that such dynamics
is rich and peculiar enough to accommodate the cuprates, or some of the
other bad actors in solid state physics.

Acknowledgement: I would like to thank Steve Strong for pointing out a
slip in an earlier version of this manuscript.

\refout

\end